\documentclass[aps,prl,twocolumn]{revtex4}
\usepackage{amsfonts,amssymb,amsmath,bm}
\usepackage{graphicx,epsfig,color}

\begin{document}

\title{Chaos and predictability of homogeneous-isotropic turbulence}
\author{G. Boffetta$^{1,2}$ and  S. Musacchio$^{3}$}

\affiliation{
$^1$Department of Physics and INFN, Universit\`a di Torino,
via P. Giuria 1, Torino, Italy \\
$^2$Institute of Atmospheric Sciences and Climate (CNR), Torino, Italy \\
$^3$Universit\'e C\^ote d'Azur, CNRS, LJAD, Nice, France
}

\begin{abstract}
We study the chaoticity and the predictability of a turbulent flow
on the basis of high-resolution direct numerical simulations at
different Reynolds numbers. 
We find that the Lyapunov exponent of turbulence, which measures
the exponential separation of two initially close solution of the 
Navier-Stokes equations, grows with the Reynolds number of the flow,
with an anomalous scaling exponent, larger than the one 
obtained on dimensional grounds.
For large perturbations, the error is transferred to larger, slower
scales where it grows algebraically generating an ``inverse cascade''
of perturbations in the inertial range. In this regime our simulations
confirm the classical predictions based on closure models of turbulence.
We show how to link chaoticity and predictability of a turbulent flow
in terms of a finite size extension of the Lyapunov exponent.
\end{abstract}


\maketitle

The strong chaoticity of turbulence does not spoil completely its
predictability. Such apparent paradox is related to the hierarchy of 
timescales in the dynamics of turbulence which ranges from the fastest 
Kolmogorov time to the slowest integral time. 

Ruelle argued many years ago
that the growth of infinitesimal 
perturbations in turbulence is ruled by the fastest timescale
\cite{ruelle1979microscopic}.
This leads to the prediction that the Lyapunov exponent 
is proportional to the inverse of the Kolmogorov time, 
and hence it increases with the Reynolds number. 
Turbulent flows at high $Re$ are therefore strongly chaotic
\cite{deissler1986navier}.
Nonetheless, the time that it takes for a small perturbation to
affect significantly the dynamics of the large scales is expected to 
be of the order of the slow integral time \cite{lorenz1969predictability}. 
The ratio between these extreme timescales increases with the Reynolds 
number and therefore allows a finite predictability time to coexist
with strong chaos \cite{boffetta2002predictability}.
This is evident from everyday experience: while the Kolmogorov time
of the atmosphere (in the planetary boundary layer) is a fraction
of a second \cite{garratt1992atmospheric} the weather is predictable 
for days.

The study of the predictability problem in turbulence dates back 
to the pioneering works of Lorenz \cite{lorenz1969predictability}
and of Leith and Kraichnan
\cite{leith1971atmospheric,leith1972predictability}.
The main idea of those studies is that a finite perturbation 
at a given scale in the inertial range of turbulence grows with the
characteristic time at that scale. 
Therefore, while an infinitesimal perturbation is expected to grow
exponentially fast, finite perturbations grow only algebraically in time, 
making the predictability of the flow much longer. 
These ideas were applied to the predictability of decaying turbulence
\cite{metais1986statistical}, two-dimensional turbulence 
\cite{kida1990error,boffetta2001predictability}
and three-dimensional turbulence at moderate Reynolds numbers
\cite{kida1992spatiotemporal}.

In this letter we investigate, on the basis of high-resolution direct 
numerical simulations, chaos in homogeneous-isotropic turbulence 
by measuring the growth of the separation between two realizations
starting from very close initial conditions. 
In the limit of infinitesimal separation we
compute the leading Lyapunov exponent of the flow 
(the rate of exponential growth of the separation \cite{ott2002chaos})
and we find that it increases with the Reynolds number, but surprisingly 
faster than what predicted on dimensional grounds \cite{ruelle1979microscopic}
and what observed in low-dimensional models of turbulence 
\cite{crisanti1993intermittency}.
For larger separation we observe the transition to an algebraic 
growth of the error, in agreement with the predictions of closure models 
\cite{leith1972predictability}.
Finally, we discuss the relation between chaoticity and the predictability
time of turbulence (defined as the average time for the perturbation
to reach a given threshold) 
in terms of the finite-size generalization of the Lyapunov exponents. 

We consider the dynamics of an incompressible velocity field 
${\bm u}({\bm x},t)$ given by the Navier-Stokes equations
\begin{equation}
\partial_t {\bm u}  + {\bm u}\cdot {\bm \nabla} {\bm u} = - \bm
{\bm \nabla} P + \nu \Delta {\bm u}  + {\bm f} \;, 
\label{eq:2}
\end{equation}
where $P$ is the pressure field and 
$\nu$ is the kinematic viscosity of the fluid. 
The term ${\bm f}$ represents a mechanical forcing needed to 
sustain the flow.
In the following we will present results in which ${\bm f}$ is a 
deterministic forcing with imposed energy input 
\cite{machiels1997predictability,lamorgese2005direct}.
The Navier-Stokes is solved numerically by a fully parallel 
pseudo-spectral code 
in a cubic box of size $\mathcal{L}$ at resolution $N^3$ 
with periodic boundary conditions in the three directions.
The main parameters of the simulations are reported in Table~\ref{table1}
and further details are found in the 
{\it Supplementary Material}.

In presence of forcing and dissipation, 
the turbulent flow reaches a statistically steady state
in which the energy dissipation rate 
$\varepsilon=\nu \langle (\partial_{\alpha} u_{\beta})^2 \rangle$ 
is equal to the
input of energy provided by the forcing 
(brackets indicate average over the physical space).
The turbulent state is characterized by a Kolmogorov energy spectrum 
$E(k) = C \varepsilon^{2/3} k^{-5/3}$. 
The kinetic energy $E=\int E(k) dk=(1/2) \langle |{\bm u}|^2 \rangle$
fluctuates around a constant mean value, which defines the typical
intensity of the large scale flow $U = (2E/3)^{1/2}$. 
The integral time is defined as $T=E/\varepsilon$
and the integral scale is $L=UT$.

We performed a series of simulations at increasing Reynolds number $Re=UL/\nu$. 
In order to ensure that the viscous range is resolved with the same
accuracy in all the simulations, the increase of $Re$ as been achieved
by increasing the resolution $N$ and reducing the viscosity 
in order to keep fixed $k_{max} \eta = 1.7$,
where $k_{max}= N/3$ is the maximum resolved wavenumber 
and $\eta = (\nu^3 /\varepsilon)^{1/4}$ is the Kolmogorov scale.

\begin{table}[ht!]
\begin{tabular}{|cccccccc|}
\hline
\hline
$N$ & $Re$ & $E$ & $U$ & $L$ & $\eta$ & $\tau_{\eta}$ & $\lambda$ \\
\hline
$1024$ & $8224$ & $0.700$ & $0.683$ & $4.78$ & $0.005$ & $0.063$ & $2.72$ \\
$512$ & $3062$ & $0.678$ & $0.672$ & $4.56$ & $0.01$ & $0.10$ & $1.39$ \\
$256$ & $1170$ & $0.665$ & $0.666$ & $4.43$ & $0.02$ & $0.16$ & $0.76$ \\
$128$ & $434$ & $0.643$ & $0.655$ & $4.21$ & $0.04$ & $0.25$ & $0.44$  \\
\hline
\hline
\end{tabular}
\caption{Parameters of the simulations. 
For all the simulations the energy input is $\varepsilon=0.1$, 
and the box size is $\mathcal{L}= 2\pi$.
$N$ is the grid resolution, $Re=UL/\nu$ the Reynolds number, $E$ the 
kinetic energy,
$U=(2E/3)^{1/2}$ is the large-scale velocity, 
$L=UE/\varepsilon$ the integral scale,
$\eta=(\nu^3/\varepsilon)^{1/4}$ the Kolmogorov scale, 
$\tau_{\eta}=(\nu/\varepsilon)^{1/2}$ the Kolmogorov time and 
$\lambda$ the Lyapunov exponent.}
\label{table1}
\end{table}

For the study of chaos and predictability we are interested in 
measuring the growth of an uncertainty in the velocity field. 
Starting from an initial velocity field ${\bm u}_1({\bm x},0)$
in the stationary turbulent state, we generate a 
perturbed velocity field ${\bm u}_2({\bm x},0)$, 
obtained by adding to the reference field a small 
white noise (the relative amplitude of the perturbation is 
$O(10^{-4})$).
We consider very small initial perturbations in order to 
guarantee that the separation between the two realizations 
is along the most unstable direction in phase space when the
error enters in the non-linear stage and therefore we do not
consider the effect of the distribution of the initial error 
on the predictability of the flow \cite{hayashi2013predictability}.
The two realizations of the velocity field are then simultaneously 
evolved in time according to (\ref{eq:2}). 
For each resolution, we performed an average over several 
independent realizations.

A natural measure of the uncertainty
is the error energy $E_\Delta(t)$ 
and the error energy spectrum $E_\Delta(k,t)$, 
defined on the basis of the error field 
$\delta {\bm u} \equiv ({\bm u}_2-{\bm u}_1)/\sqrt{2}$ as
\begin{equation}
E_{\Delta}(t) = \int_0^{\infty} E_{\Delta}(k,t) dk
= {1 \over 2} \langle |\delta {\bm u}({\bm x},t)|^2 \rangle \;.
\label{eq:3}
\end{equation}
With the normalization coefficient $1/\sqrt{2}$ we have
$E_{\Delta}=E$ for completely uncorrelated fields.

\begin{figure}
\centering\includegraphics[width=0.47\textwidth]{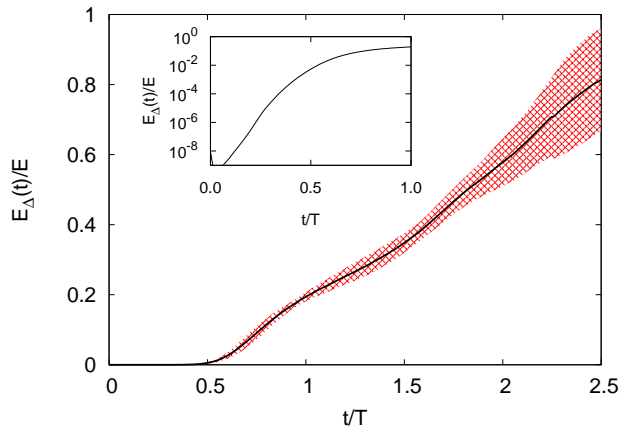}
\caption{(Color online) Error energy $E_\Delta(t)$ growth for the simulation
at $N=1024$.
The error energy is averaged over $10$ different realizations (black line). 
The fluctuations of the error energy within one standard deviation
from the mean are represented by the shaded area.  
Inset: The initial exponential growth of the error.}
\label{fig1}
\end{figure}

Figure~\ref{fig1} shows the time evolution of the error energy
$E_\Delta$ for the simulation at the highest $Re$,
averaged over an ensemble of $10$ independent
realizations. 
In the initial stage the error grows exponentially as 
$E_\Delta(t) = E_\Delta(0) \exp(L_2 t)$ (see inset of Fig.~\ref{fig1})
where $L_2$ is the generalized Lyapunov exponent of order
$2$~\cite{cencini2010chaos}.
At later times we observe a regime of linear growth of the error 
$E_\Delta(t) \simeq \varepsilon t$. 
The growth rate displays large fluctuations
as the error approaches its saturation value $E_\Delta(t) \simeq E$.
This is due to the fluctuations of the kinetic energy which occur
on the same time scale of the saturation of the error and are
associated to the dynamics of the large scales.
It is worth to notice that the late regime of saturation of the error 
might display a non-universal behavior with respect to the forcing mechanism. 
As an example, the deterministic force used in our study 
is proportional to the large-scale velocity. 
At late times, when the error has significantly affected the large scales, 
the force acting on the two fields ${\bm u}_1$ and ${\bm u}_2$ becomes 
different. 
This could induce a faster saturation of the error 
with respect to other forcing mechanism which enforce large-scale correlations.

\begin{figure}
\centering\includegraphics[width=0.47\textwidth]{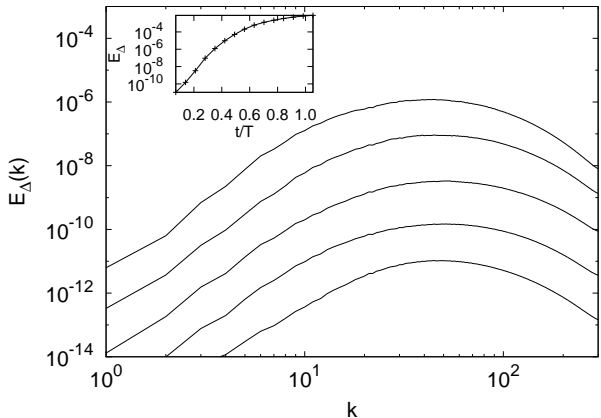}
\caption{The spectrum of the error $E_\Delta(k,t)$ at times 
$t/T=0.07, 0.14, 0.21, 0.28, 0.35$ (from bottom to top) 
in the linear phase for the simulation at $N=1024$ averaged
over $10$ independent realizations.
Inset: The error energy $E_{\Delta}$ as a function of time 
in semilogarithmic plot.}
\label{fig2}
\end{figure}

During the initial stage of exponential growth 
the error energy spectrum $E_\Delta(k,t)$ 
is peaked at wavenumbers around the dissipation range 
$k \simeq k_\eta \simeq 1/\eta$ and grows exponentially 
in a self similar way, as shown in Fig.~\ref{fig2}.

\begin{figure}
\centering\includegraphics[width=0.47\textwidth]{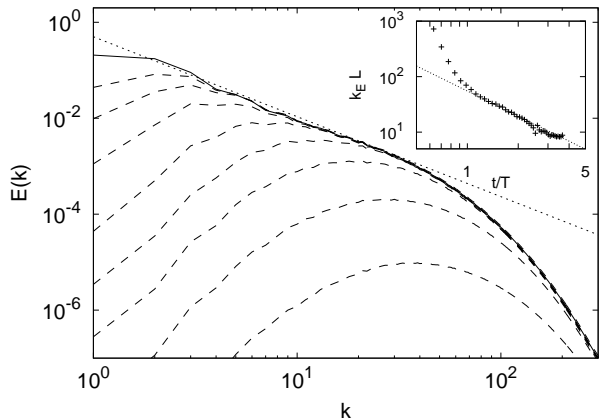}
\caption{The spectrum of the error $E_\Delta(k,t)$ at times 
$t/T=0.42, 0.56, 0.70, 0.84, 1.1, 1.4, 1.8, 2.1$ 
(dashed lines, from bottom to top) 
compared with the stationary energy spectrum $E(k)$ (solid line) 
for simulations at $N=1024$ averaged
over $10$ independent realizations.
The dotted line represents the Kolmogorov scaling $k^{-5/3}$. 
Inset: The error wavenumber $k_E$ as a function of time (crosses),
compared with the dimensional scaling $k_E \sim t^{3/2}$ (dotted line).}
\label{fig3}
\end{figure}

At later times, the error propagates to lower wavenumbers
and the error spectrum develops
a scaling range $E_\Delta(k) \sim k^{-5/3}$ (see Fig.~\ref{fig3}).
At each time it is possible to identify the error wavenumber $k_E(t)$ 
at which the error energy spectrum has reached a given fraction 
$\alpha \simeq 1$
of the energy spectrum $E_\Delta(k_E,t) /E(k_E) = \alpha$. 
The two velocity fields ${\bm u}_1$ and ${\bm u}_2$ can be then assumed to
be completely decorrelated at scales smaller than $1/k_E$ 
and still correlated at larger scales.  

The transition from the exponential growth to the linear growth of $E_\Delta$
occurs when the two fields are completely decorrelated on the
dissipative scales, that is when $k_E \simeq k_\eta$ . 
The idea, originally proposed by Lorenz \cite{lorenz1969predictability},
is that the time that it takes to decorrelate completely the two
fields at a given scale $\ell \simeq 1/k$ within the inertial range 
is proportional to the turnover time of the eddies at that scale
$\tau_\ell \sim \varepsilon^{-1/3} \ell^{2/3}$ \cite{frisch1995turbulence}.
This leads to the dimensional prediction 
\begin{equation}
k_E(t) \simeq \varepsilon^{-1/2} t^{-3/2} 
\label{eq:4}
\end{equation}
for the evolution of the error wavenumber, 
which is confirmed by our numerical finding (see inset of
Fig.~\ref{fig2}). 

Equation~(\ref{eq:4}) provides an estimation of the 
predictability time $T_P$ that an infinitesimal 
error takes to contaminate a given wavenumber $k$,
$T_p(k) = A \varepsilon^{-1/3} k^{-2/3}$
\cite{leith1972predictability,cardesa2015temporal}
where the dimensionless coefficient 
$A$ depends on the threshold $\alpha$ (and possibly on the Reynolds
number).
In our simulation at $Re=8516$ we measure $A=12$ for $\alpha=0.5$ 
to be compared with the value $A=10$ obtained from early studies
with closure models 
in the limit of infinite $Re$~\cite{leith1972predictability}.  

Integrating the error spectrum with the ansatz
$E_\Delta(k,t) = 0$ for $k < k_E(t)$  
$E_\Delta(k,t) = E(k)$ for $k > k_E(t)$ 
and using the dimensional scaling (\ref{eq:4}), 
one obtains the prediction for the linear growth of the error energy: 
\begin{equation}
E_\Delta(t) = G \varepsilon t \;.
\label{eq:5} 
\end{equation} 
The value of the dimensionless constant $G$ measured in the simulation at 
$Re=8516$ is $G=0.45 \pm 0.05$, not far from that obtained by
the test field model closure $G=0.23$ \cite{leith1972predictability}.


As already discussed, in the early stage the perturbation can 
be considered infinitesimal
and therefore grows exponentially as shown in the inset of Fig.~\ref{fig1}.
This is the signature of the chaotic nature of the flow and the 
predictability is characterized by the Lyapunov exponent $\lambda$.
On dimensional grounds the Lyapunov exponent 
can be assumed to be proportional to the inverse of the fastest 
time-scale of the flow, i.e., the Kolmogorov timescale
$\tau_{\eta}=(\nu/\varepsilon)^{1/2}$~\cite{ruelle1979microscopic}. 
Since the ratio between $\tau_\eta$ and the integral timescale $T$ 
increases with the Reynolds number as $T/\tau_{\eta} \sim Re^{1/2}$ 
one has the prediction that the Lyapunov exponent is proportional to the 
square root of the Reynolds number: 
\begin{equation}
\lambda \simeq \tau_{\eta}^{-1} \simeq T^{-1} Re^{1/2} \;.
\label{eq:6} 
\end{equation} 
Therefore the predictability time $T_P$ for
infinitesimal perturbations vanishes in the limit of large $Re$. 

The dimensional prediction (\ref{eq:6}) is obtained under the
assumption of self-similarity of the velocity field with 
Kolmogorov scaling exponent $h=1/3$ \cite{frisch1995turbulence}
For a generic exponent $h \in (0:1)$ one has 
$\lambda \simeq \tau_{\eta}^{-1} \simeq T^{-1} Re^{\beta}$
with $\beta = (1-h)/(1+h)$. 
Averaging over the multifractal spectrum $\mathcal{D}(h)$ 
allows to take into account intermittency
corrections and this gives $\beta = 0.459$ 
\cite{crisanti1993intermittency,aurell1996growth}.

\begin{figure}
\centering\includegraphics[width=0.47\textwidth]{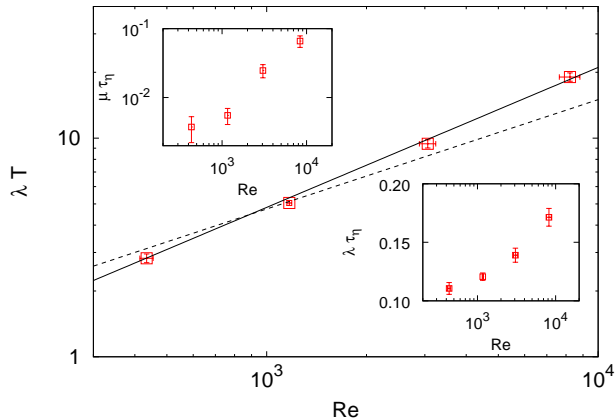}
\caption{(Color online) Lyapunov exponents $\lambda$ as a function of
$Re$ (squares). The solid line represents the best fit scaling 
$\lambda T \simeq Re^{0.64}$ while the dashed line is the dimensional scaling 
$\lambda T \simeq Re^{1/2}$. 
Lower inset: The Lyapunov exponents $\lambda$ compensated with the Kolmogorov
time scale $\tau_{\eta}$ as a function of $Re$.
Upper inset: The Lyapunov variance $\mu$ compensated with the Kolmogorov
time scale $\tau_{\eta}$ as a function of $Re$.}
\label{fig4}
\end{figure}

We have computed the Lyapunov exponent $\lambda$ 
by measuring the average rate of logarithmic divergence of two
close realizations, a standard method in the study of dynamical systems
\cite{benettin1976kolmogorov,benettin1980lyapunov,cencini2010chaos},
for the simulations at different Reynolds numbers 
(see Table~\ref{table1}).
Interestingly, we find that the Lyapunov exponent increases 
with $Re$ faster than the dimensional prediction (\ref{eq:6}), 
as shown in Fig.~\ref{fig4}.
Fitting the measured values with a power law $\lambda T \simeq
Re^{\beta}$ gives the exponent $\beta = 0.64 \pm 0.05$. 
It is remarkable that the measured deviation from the 
dimensional prediction $\beta=0.5$ is opposite with respect to the 
predicted correction due to intermittency.
Our findings suggest that the dimensional estimate of the Lyapunov
exponent as the inverse Kolmogorov time does not give an accurate 
characterization of the chaoticity of a turbulent flow. 
The inset of Fig.~\ref{fig4} shows that indeed the quantity
$\lambda \tau_{\eta}$ increases with $Re$.

Since the Lyapunov exponent is an average quantity, it is interesting
to investigate its fluctuations and their dependence on $Re$. We 
have therefore measured the variance $\mu$ of the distribution 
of the finite-time Lyapunov exponents,
a standard measure of the fluctuations in a chaotic system
\cite{ott2002chaos,cencini2010chaos} (see also the 
{\it Supplementary Material}).
The results, plotted in Fig.~\ref{fig4}, shows that also 
$\mu \tau_{\eta}$ increases with $Re$ and faster than the 
Lyapunov exponent (a fit gives $\mu T \simeq Re^{1.2}$ although the 
errors here are large). 

The connection between predictability and chaoticity in 
turbulent flows can be extended also to finite perturbations,
of the order of the velocities of the inertial range, by means of 
the finite size Lyapunov exponents (FSLE) $\Lambda (\delta)$. 
The FSLE has been introduced to  measure the chaoticity of systems
with many characteristic time scales
\cite{aurell1996growth,boffetta2002predictability}.
It is defined in terms of the average time $T_r(\delta)$ 
that it takes for a perturbation of size $\delta$ to grow by a factor $r$, as 
$\Lambda(\delta) = \ln(r)/\langle T_r(\delta)\rangle$ 
(where the average is now over different realizations).
We remind that performing averages at fixed times 
is not equivalent to averaging at fixed error
size. The latter procedure was found to be more effective in
intermittent systems, in which scaling laws can be affected by 
strong fluctuations of the error (as in Fig.~\ref{fig1}).

In the limit $\delta \to 0$ the FSLE recovers,
by definition, the usual Lyapunov exponent, i.e.
$\lim_{\delta \to 0} \Lambda(\delta) = \lambda$ \cite{aurell1996growth}.
For finite errors, $\Lambda(\delta)$ measures the average growth rate 
of the uncertainty of size $\delta$. 
Following the idea of Lorenz \cite{lorenz1969predictability} 
that a perturbation of size 
$\delta\sim u_\ell$ within the inertial range of turbulence
grows with the local eddy turnover time 
$\tau_\ell \sim \varepsilon^{-1/3} \ell^{2/3} \sim  \varepsilon^{-1}
u_\ell^2$, one obtains the prediction \cite{aurell1996growth}
\begin{equation} 
\Lambda(\delta) \simeq  \varepsilon \delta^{-2} \;.
\label{eq:7} 
\end{equation}

\begin{figure}
\centering\includegraphics[width=0.47\textwidth]{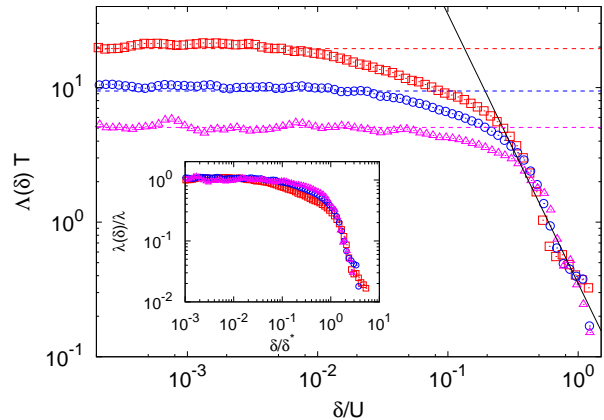}
\caption{(Color online) Finite-size Lyapunov exponents $\Lambda(\delta)$ (FSLE)
as a function of the velocity uncertainty $\delta$ for 
$N=1024$ (red squares) 
$N=512$ (blue circles) 
$N=256$ (purple triangles). 
The values of the Lyapunov exponents $\lambda$ are also shown (dashed lines).
Black solid line represents the scaling $\Lambda(\delta) \sim \delta^{-2}$. 
Inset: The FSLE $\Lambda(\delta)$
rescaled by the Lyapunov exponents $\lambda$ 
as a function of the rescaled uncertainty $\delta / \delta^*$.}
\label{fig5}
\end{figure}

In Figure~\ref{fig5} we show the FSLE as a function of the error
$\delta$ for three values of $Re$. 
For small $\delta$ the FSLE approaches the constant value 
$\Lambda(\delta) \simeq \lambda$, while in the inertial range we observe
the dimensional scaling (\ref{eq:7}). 
The crossover between the two regimes is expected to occur at 
$\delta^* \simeq (\varepsilon/\lambda)^{1/2}$.   
Rescaling the error $\delta$ with $\delta^*$ and $\Lambda(\delta)$
with $\lambda$ we find a good collapse of the two regimes of
infinitesimal and finite errors, as shown in the inset of Fig.~\ref{fig5}.
Figure~\ref{fig5} also shows that the crossover range between the two regimes 
increases with $Re$. 
One possible explanation for this long crossover 
is that the transition between the two regimes 
involves the dynamics of eddies which are at the border between the inertial 
and the dissipative scales, in the so-called 
intermediate dissipative range \cite{frisch1995turbulence}. 
The extension of this range is known to grow with the Reynolds number, 
and this could cause the broadening of the crossover regime for the FSLE.  

Remarkably, Figure~\ref{fig5} shows that in the scaling range 
$\Lambda(\delta) \sim \delta^{-2}$ 
the error growth rate $\Lambda$ becomes independent 
both on the Reynolds number and on the values of the Lyapunov exponent. 
The independence of the FSLE in the scaling range on the value $\lambda$ 
observed for infinitesimal errors provides a
clear explanation of how in turbulent flows it is possible to observe 
the coexistence of long predictability time
at large scales and strong chaoticity at small scales.

In conclusion, we studied the chaotic and predictability properties 
of fully developed 
turbulence by simulating two realizations of the velocity field initially
separated by a very small perturbation. 
At short times the separation increases exponentially as a consequence
of the chaoticity of the flow. 
Finite perturbations increase linearly in time, as predicted by dimensional 
arguments, and the time for the perturbation
to affect a wavenumber $k$ in the inertial range is proportional
to $\varepsilon^{-1/3} k^{-2/3}$.

The Lyapunov exponent is found to grow with the Reynolds number faster
than what predicted by a dimensional argument and intermittency models
and, as a consequence, the product $\lambda \tau_{\eta}$ grows with $Re$.
This indicates that the strong, intermittent fluctuations of turbulence
at small scales give diverse contributions on different observables.
In addition to the interest for many applications, turbulence
is a prototypical example of system with many scales and characteristic times. 
Our results on the chaoticity of turbulence and
its dependence on the number of active degrees of freedom are
therefore of general interest for the study of extended dynamical
systems.


\acknowledgments
The Authors gratefully acknowledge support from the Simons Center for 
Geometry and Physics, Stony Brook University, where part of this
work was performed. The COST Action MP1305, supported by COST 
(European Cooperation in Science and Technology) is acknowledged.
Numerical simulations have been performed at Cineca within the INFN-Cineca
agreement {\it INF17\_fldturb}.

\bibliography{biblio}

\begin{thebibliography}{22}
\expandafter\ifx\csname natexlab\endcsname\relax\def\natexlab#1{#1}\fi
\expandafter\ifx\csname bibnamefont\endcsname\relax
  \def\bibnamefont#1{#1}\fi
\expandafter\ifx\csname bibfnamefont\endcsname\relax
  \def\bibfnamefont#1{#1}\fi
\expandafter\ifx\csname citenamefont\endcsname\relax
  \def\citenamefont#1{#1}\fi
\expandafter\ifx\csname url\endcsname\relax
  \def\url#1{\texttt{#1}}\fi
\expandafter\ifx\csname urlprefix\endcsname\relax\def\urlprefix{URL }\fi
\providecommand{\bibinfo}[2]{#2}
\providecommand{\eprint}[2][]{\url{#2}}

\bibitem[{\citenamefont{Ruelle}(1979)}]{ruelle1979microscopic}
\bibinfo{author}{\bibfnamefont{D.}~\bibnamefont{Ruelle}},
  \bibinfo{journal}{Phys. Lett.} \textbf{\bibinfo{volume}{72A}},
  \bibinfo{pages}{81} (\bibinfo{year}{1979}).

\bibitem[{\citenamefont{Deissler}(1986)}]{deissler1986navier}
\bibinfo{author}{\bibfnamefont{R.~G.} \bibnamefont{Deissler}},
  \bibinfo{journal}{Phys. Fluids} \textbf{\bibinfo{volume}{29}},
  \bibinfo{pages}{1453} (\bibinfo{year}{1986}).

\bibitem[{\citenamefont{Lorenz}(1969)}]{lorenz1969predictability}
\bibinfo{author}{\bibfnamefont{E.~N.} \bibnamefont{Lorenz}},
  \bibinfo{journal}{Tellus} \textbf{\bibinfo{volume}{21}}, \bibinfo{pages}{289}
  (\bibinfo{year}{1969}).

\bibitem[{\citenamefont{Boffetta et~al.}(2002)\citenamefont{Boffetta, Cencini,
  Falcioni, and Vulpiani}}]{boffetta2002predictability}
\bibinfo{author}{\bibfnamefont{G.}~\bibnamefont{Boffetta}},
  \bibinfo{author}{\bibfnamefont{M.}~\bibnamefont{Cencini}},
  \bibinfo{author}{\bibfnamefont{M.}~\bibnamefont{Falcioni}}, \bibnamefont{and}
  \bibinfo{author}{\bibfnamefont{A.}~\bibnamefont{Vulpiani}},
  \bibinfo{journal}{Phys. Rep.} \textbf{\bibinfo{volume}{356}},
  \bibinfo{pages}{367} (\bibinfo{year}{2002}).

\bibitem[{\citenamefont{Garratt}(1992)}]{garratt1992atmospheric}
\bibinfo{author}{\bibfnamefont{J.~R.} \bibnamefont{Garratt}},
  \emph{\bibinfo{title}{The atmospheric boundary layer}}, vol.
  \bibinfo{volume}{416} (\bibinfo{publisher}{Cambridge University Press},
  \bibinfo{year}{1992}).

\bibitem[{\citenamefont{Leith}(1971)}]{leith1971atmospheric}
\bibinfo{author}{\bibfnamefont{C.~E.} \bibnamefont{Leith}},
  \bibinfo{journal}{J. Atmos. Sci.} \textbf{\bibinfo{volume}{28}},
  \bibinfo{pages}{145} (\bibinfo{year}{1971}).

\bibitem[{\citenamefont{Leith and Kraichnan}(1972)}]{leith1972predictability}
\bibinfo{author}{\bibfnamefont{C.~E.} \bibnamefont{Leith}} \bibnamefont{and}
  \bibinfo{author}{\bibfnamefont{R.~H.} \bibnamefont{Kraichnan}},
  \bibinfo{journal}{J. Atmos. Sci.} \textbf{\bibinfo{volume}{29}},
  \bibinfo{pages}{1041} (\bibinfo{year}{1972}).

\bibitem[{\citenamefont{M{\'e}tais and Lesieur}(1986)}]{metais1986statistical}
\bibinfo{author}{\bibfnamefont{O.}~\bibnamefont{M{\'e}tais}} \bibnamefont{and}
  \bibinfo{author}{\bibfnamefont{M.}~\bibnamefont{Lesieur}},
  \bibinfo{journal}{J. Atmos. Sci.} \textbf{\bibinfo{volume}{43}},
  \bibinfo{pages}{857} (\bibinfo{year}{1986}).

\bibitem[{\citenamefont{Kida et~al.}(1990)\citenamefont{Kida, Yamada, and
  Ohkitani}}]{kida1990error}
\bibinfo{author}{\bibfnamefont{S.}~\bibnamefont{Kida}},
  \bibinfo{author}{\bibfnamefont{M.}~\bibnamefont{Yamada}}, \bibnamefont{and}
  \bibinfo{author}{\bibfnamefont{K.}~\bibnamefont{Ohkitani}},
  \bibinfo{journal}{J. Phys. Soc. Japan} \textbf{\bibinfo{volume}{59}},
  \bibinfo{pages}{90} (\bibinfo{year}{1990}).

\bibitem[{\citenamefont{Boffetta and
  Musacchio}(2001)}]{boffetta2001predictability}
\bibinfo{author}{\bibfnamefont{G.}~\bibnamefont{Boffetta}} \bibnamefont{and}
  \bibinfo{author}{\bibfnamefont{S.}~\bibnamefont{Musacchio}},
  \bibinfo{journal}{Phys. Fluids} \textbf{\bibinfo{volume}{13}},
  \bibinfo{pages}{1060} (\bibinfo{year}{2001}).

\bibitem[{\citenamefont{Kida and Ohkitani}(1992)}]{kida1992spatiotemporal}
\bibinfo{author}{\bibfnamefont{S.}~\bibnamefont{Kida}} \bibnamefont{and}
  \bibinfo{author}{\bibfnamefont{K.}~\bibnamefont{Ohkitani}},
  \bibinfo{journal}{Phys. Fluids A} \textbf{\bibinfo{volume}{4}},
  \bibinfo{pages}{1018} (\bibinfo{year}{1992}).

\bibitem[{\citenamefont{Ott}(2002)}]{ott2002chaos}
\bibinfo{author}{\bibfnamefont{E.}~\bibnamefont{Ott}},
  \emph{\bibinfo{title}{Chaos in dynamical systems}}
  (\bibinfo{publisher}{Cambridge University Press}, \bibinfo{year}{2002}).

\bibitem[{\citenamefont{Crisanti et~al.}(1993)\citenamefont{Crisanti, Jensen,
  Vulpiani, and Paladin}}]{crisanti1993intermittency}
\bibinfo{author}{\bibfnamefont{A.}~\bibnamefont{Crisanti}},
  \bibinfo{author}{\bibfnamefont{M.~H.} \bibnamefont{Jensen}},
  \bibinfo{author}{\bibfnamefont{A.}~\bibnamefont{Vulpiani}}, \bibnamefont{and}
  \bibinfo{author}{\bibfnamefont{G.}~\bibnamefont{Paladin}},
  \bibinfo{journal}{Phys. Rev. Lett.} \textbf{\bibinfo{volume}{70}},
  \bibinfo{pages}{166} (\bibinfo{year}{1993}).

\bibitem[{\citenamefont{Machiels}(1997)}]{machiels1997predictability}
\bibinfo{author}{\bibfnamefont{L.}~\bibnamefont{Machiels}},
  \bibinfo{journal}{Phys. Rev. Lett.} \textbf{\bibinfo{volume}{79}},
  \bibinfo{pages}{3411} (\bibinfo{year}{1997}).

\bibitem[{\citenamefont{Lamorgese et~al.}(2005)\citenamefont{Lamorgese,
  Caughey, and Pope}}]{lamorgese2005direct}
\bibinfo{author}{\bibfnamefont{A.~G.} \bibnamefont{Lamorgese}},
  \bibinfo{author}{\bibfnamefont{D.~A.} \bibnamefont{Caughey}},
  \bibnamefont{and} \bibinfo{author}{\bibfnamefont{S.~B.} \bibnamefont{Pope}},
  \bibinfo{journal}{Phys. Fluids} \textbf{\bibinfo{volume}{17}},
  \bibinfo{pages}{015106} (\bibinfo{year}{2005}).

\bibitem[{\citenamefont{Hayashi et~al.}(2013)\citenamefont{Hayashi, Ishihara,
  and Kaneda}}]{hayashi2013predictability}
\bibinfo{author}{\bibfnamefont{K.}~\bibnamefont{Hayashi}},
  \bibinfo{author}{\bibfnamefont{T.}~\bibnamefont{Ishihara}}, \bibnamefont{and}
  \bibinfo{author}{\bibfnamefont{Y.}~\bibnamefont{Kaneda}},
  \bibinfo{journal}{Statistical Theories and Computational Approaches to
  Turbulence} p. \bibinfo{pages}{239} (\bibinfo{year}{2013}).

\bibitem[{\citenamefont{Cencini et~al.}(2010)\citenamefont{Cencini, Cecconi,
  and Vulpiani}}]{cencini2010chaos}
\bibinfo{author}{\bibfnamefont{M.}~\bibnamefont{Cencini}},
  \bibinfo{author}{\bibfnamefont{F.}~\bibnamefont{Cecconi}}, \bibnamefont{and}
  \bibinfo{author}{\bibfnamefont{A.}~\bibnamefont{Vulpiani}},
  \emph{\bibinfo{title}{Chaos: from simple models to complex systems}},
  vol.~\bibinfo{volume}{17} (\bibinfo{publisher}{World Scientific},
  \bibinfo{year}{2010}).

\bibitem[{\citenamefont{Frisch}(1995)}]{frisch1995turbulence}
\bibinfo{author}{\bibfnamefont{U.}~\bibnamefont{Frisch}},
  \emph{\bibinfo{title}{Turbulence}} (\bibinfo{publisher}{Cambridge Univ.
  Press}, \bibinfo{year}{1995}).

\bibitem[{\citenamefont{Cardesa et~al.}(2015)\citenamefont{Cardesa,
  Vela-Mart{\'\i}n, Dong, and Jim{\'e}nez}}]{cardesa2015temporal}
\bibinfo{author}{\bibfnamefont{J.~I.} \bibnamefont{Cardesa}},
  \bibinfo{author}{\bibfnamefont{A.}~\bibnamefont{Vela-Mart{\'\i}n}},
  \bibinfo{author}{\bibfnamefont{S.}~\bibnamefont{Dong}}, \bibnamefont{and}
  \bibinfo{author}{\bibfnamefont{J.}~\bibnamefont{Jim{\'e}nez}},
  \bibinfo{journal}{Phys. Fluids} \textbf{\bibinfo{volume}{27}},
  \bibinfo{pages}{111702} (\bibinfo{year}{2015}).

\bibitem[{\citenamefont{Aurell et~al.}(1996)\citenamefont{Aurell, Boffetta,
  Crisanti, Paladin, and Vulpiani}}]{aurell1996growth}
\bibinfo{author}{\bibfnamefont{E.}~\bibnamefont{Aurell}},
  \bibinfo{author}{\bibfnamefont{G.}~\bibnamefont{Boffetta}},
  \bibinfo{author}{\bibfnamefont{A.}~\bibnamefont{Crisanti}},
  \bibinfo{author}{\bibfnamefont{G.}~\bibnamefont{Paladin}}, \bibnamefont{and}
  \bibinfo{author}{\bibfnamefont{A.}~\bibnamefont{Vulpiani}},
  \bibinfo{journal}{Phys. Rev. Lett.} \textbf{\bibinfo{volume}{77}},
  \bibinfo{pages}{1262} (\bibinfo{year}{1996}).

\bibitem[{\citenamefont{Benettin et~al.}(1976)\citenamefont{Benettin, Galgani,
  and Strelcyn}}]{benettin1976kolmogorov}
\bibinfo{author}{\bibfnamefont{G.}~\bibnamefont{Benettin}},
  \bibinfo{author}{\bibfnamefont{L.}~\bibnamefont{Galgani}}, \bibnamefont{and}
  \bibinfo{author}{\bibfnamefont{J.}~\bibnamefont{Strelcyn}},
  \bibinfo{journal}{Phys. Rev. A} \textbf{\bibinfo{volume}{14}},
  \bibinfo{pages}{2338} (\bibinfo{year}{1976}).

\bibitem[{\citenamefont{Benettin et~al.}(1980)\citenamefont{Benettin, Galgani,
  Giorgilli, and Strelcyn}}]{benettin1980lyapunov}
\bibinfo{author}{\bibfnamefont{G.}~\bibnamefont{Benettin}},
  \bibinfo{author}{\bibfnamefont{L.}~\bibnamefont{Galgani}},
  \bibinfo{author}{\bibfnamefont{A.}~\bibnamefont{Giorgilli}},
  \bibnamefont{and} \bibinfo{author}{\bibfnamefont{J.}~\bibnamefont{Strelcyn}},
  \bibinfo{journal}{Meccanica} \textbf{\bibinfo{volume}{15}},
  \bibinfo{pages}{9} (\bibinfo{year}{1980}).

\end{thebibliography}

\end{document}